\documentclass[aps,prx,twocolumn,longbibliography,superscriptaddress,noeprint,reprint]{revtex4-2}
\usepackage{amsmath}
\usepackage{graphicx}
\usepackage{mathtools}
\usepackage{tikz}
\usepackage{siunitx}
\usepackage{xcolor}
\usepackage[normalem]{ulem}
\usepackage{hyperref}
\usepackage{wrapfig}
\usepackage{xr}
\usepackage[english]{babel}

\usepackage{float}
\usepackage{multirow}
\usepackage{sidecap}

\usepackage{xcolor}
\hypersetup{
    colorlinks,
    linkcolor={red!50!black},
    citecolor={blue!50!black},
    urlcolor={blue!80!black}
}

\usepackage{color}
\definecolor{blue}{rgb}{0.00,0.00,0.95}

\renewcommand{\d}{{\rm d}}

\begin{document}

\title{Learning the interactions underlying multicellular migration using many-body stochastic inference}

\author{Bram J. Hoogland}
 \thanks{These authors contributed equally\vspace{-0.3em}}
\affiliation{Department of Physics and Astronomy, Vrije Universiteit Amsterdam, De Boelelaan 1105, 1081 HV Amsterdam, The Netherlands}
\author{Tom Brandst\"atter}
 \thanks{These authors contributed equally\vspace{-0.3em}}
\affiliation{Department of Physics and Astronomy, Vrije Universiteit Amsterdam, De Boelelaan 1105, 1081 HV Amsterdam, The Netherlands}
\affiliation{Arnold-Sommerfeld-Center for Theoretical Physics, Ludwig-Maximilians-Universität München, Theresienstr. 37, 80333 Munich, Germany}
\author{Emily Brieger}
\affiliation{Faculty of Physics and Center for NanoScience, Ludwig-Maximilians-Universität München, Geschwister-Scholl-Platz 1, 80539 Munich, Germany}
\author{Joachim O. R\"adler}
\affiliation{Faculty of Physics and Center for NanoScience, Ludwig-Maximilians-Universität München, Geschwister-Scholl-Platz 1, 80539 Munich, Germany}
\author{Chase P. Broedersz}
 \thanks{Corresponding author: c.p.broedersz@vu.nl \vspace{-1.3em}}
\affiliation{Department of Physics and Astronomy, Vrije Universiteit Amsterdam, De Boelelaan 1105, 1081 HV Amsterdam, The Netherlands}

\begin{abstract}

Uncovering the interactions that give rise to emergent collective behavior in active and living matter remains a major challenge. Multicellular migration exemplifies this challenge: complex cell–cell interactions shape collective motion, but are difficult to measure directly. Here, we introduce a many-body stochastic inference framework that learns these interactions directly from experimental trajectories. We infer interaction models spanning different dynamical dependencies and complexities from short-time cell dynamics, and select among them based on their ability to predict emergent collective behaviors. Applied across three behaviorally distinct cell types, this approach yields minimal, predictive models of cell–cell interactions underlying multicellular migration. We find that non-cancerous epithelial cells coordinate multicellular migration through reciprocal repulsion and velocity alignment, while cancerous mesenchymal cells additionally exhibit previously unrecognized nonreciprocal interactions that couple to self-propulsion through two distinct mechanisms, which we term velocity inhibition in fibrosarcoma cells and velocity enhancement in breast cancer cells. Conceptually, these nonreciprocal interactions allow cells to tune the onset and speed of flocks, underscoring their role in collective migration and potentially enhancing cancer cell invasiveness at large cell densities. More broadly, our framework offers a general and principled route for extracting interaction laws of active and living systems directly from data.
\end{abstract}

\maketitle
Collective behavior shapes the dynamics of living systems on all scales, from biomolecules that organize the interior of cells \cite{holzbaur2010coordination, guerin2010coordination}, to migrating cells that form tissues \cite{rorth2009collective, cetera2014epithelial,krndija2019active}, and to animals moving in groups \cite{couzin2005effective,ballerini2008interaction}. Although the resulting behaviors are diverse and complex, such coordination typically emerges from local interactions rather than central control. Beyond their interactions, the motion of individual agents in a collective is determined by self-propelled motility, biological variability, and inherent stochasticity \cite{romanczuk2012active}. Thus, a system's many-body stochastic dynamics arises from multiple intertwined factors acting simultaneously, making it difficult to unravel the underlying interaction rules \cite{vicsek2012collective, marchetti2013hydrodynamics}. This challenge is further compounded by the potentially nonequilibrium and complex nature of interactions in living systems, shaped by internal dynamics and environmental cues \cite{couzin2005effective,ballerini2008interaction,roycroft2016molecular, cetera2014epithelial, brandstatter2023curvature,vicsek2012collective}. Data-driven theoretical approaches have recently emerged as powerful methods to provide insight into interactive behavior \cite{brunton2016discovering,bruckner2021learning,lachance2022learning,brandstatter2025data,bialek2012statistical,katz2011inferring, yang2024learning}. Yet, learning physically interpretable dynamical interaction rules of a living many-body system from experiments that contain interactions, inherent stochasticity, and measurement noise remains an open problem. \\

Collective cell migration is an important and relevant setting for addressing this challenge \cite{rorth2009collective, friedl2009collective, haeger2015collective, alert2020physical}. As fundamental units of life, cells act as autonomous agents that exhibit functional collective behaviors, such as morphogenesis \cite{aman2010cell}, wound healing \cite{poujade2007collective}, and cancer invasion \cite{friedl2009collective}. At the heart of these cellular behaviors are interactions that integrate mechanical forces and biochemical signals \cite{ladoux2017mechanobiology,roycroft2016molecular}. For example, during collective cell migration in epithelial tissues, the coupling of cells via E-cadherin junctions is essential to establish their coherent collective behavior \cite{lecuit2015cadherin, jain2020role}. In contrast, the lack of E-cadherin-mediated interactions is associated with reduced large-scale order and cancer progression \cite{theveneau2013collective,wong2014collective,Mendonsa2018}. Although many details of the biomolecular machinery underlying these collective behaviors are becoming increasingly clear \cite{ladoux2017mechanobiology,roycroft2016molecular}, the system-level dynamical interaction rules that cells obey to create spatiotemporal self-organization at the tissue level are not well understood \cite{alert2020physical}. Despite advances in measuring traction forces and stresses in multicellular systems \cite{trepat2009physical, ladoux2017mechanobiology, gomez2020measuring, rossetti2024optogenetic}, directly measuring the interactions through which cells influence each other’s movement remains elusive.

Here, we develop a statistical inference framework to learn quantitative, cell-level dynamical interaction rules directly from multicellular migration  experiments. Rather than inferring interactions from isolated cell pairs \cite{bruckner2021learning, brandstatter2025data}, our approach learns interactions directly from the collective, where each cell can respond simultaneously to multiple neighbors. To this end, our inference pipeline searches a broad, systematically constructed space of candidate interaction laws, infers these laws directly from the cell trajectories, and discriminates between models by comparing their predicted collective statistics with experiments across densities~\cite{Vedel2013, 10.1098/rsif.2020.0143,angelini2011glass,castro2016clustering,Garcia2015d}. This allows us to identify the simplest interaction law supported by the experimental data. We apply this approach to multicellular migration on micropatterns \cite{jain2020role, bruckner2021learning, brandstatter2025data, bertrand2024clustering} with a range of behaviorally distinct human cell types \cite{theveneau2013collective, wong2014collective,lee2021distinct}. The interactions we uncover differ across cell types and, in some cases, reveal the active nature of the system through broken action–reaction symmetry \cite{bowick2022symmetry}, thereby granting cells the ability to control a wider range of collective behaviors \cite{d2017contact,hayakawa2020polar,fruchart2021non, copenhagen2018frustration}. Our statistical inference framework offers a broadly applicable method to learn the stochastic many-body dynamical laws of active and living matter from experiments.

\section*{Results}
\begin{figure*}[t!]
	\includegraphics[width=\textwidth]{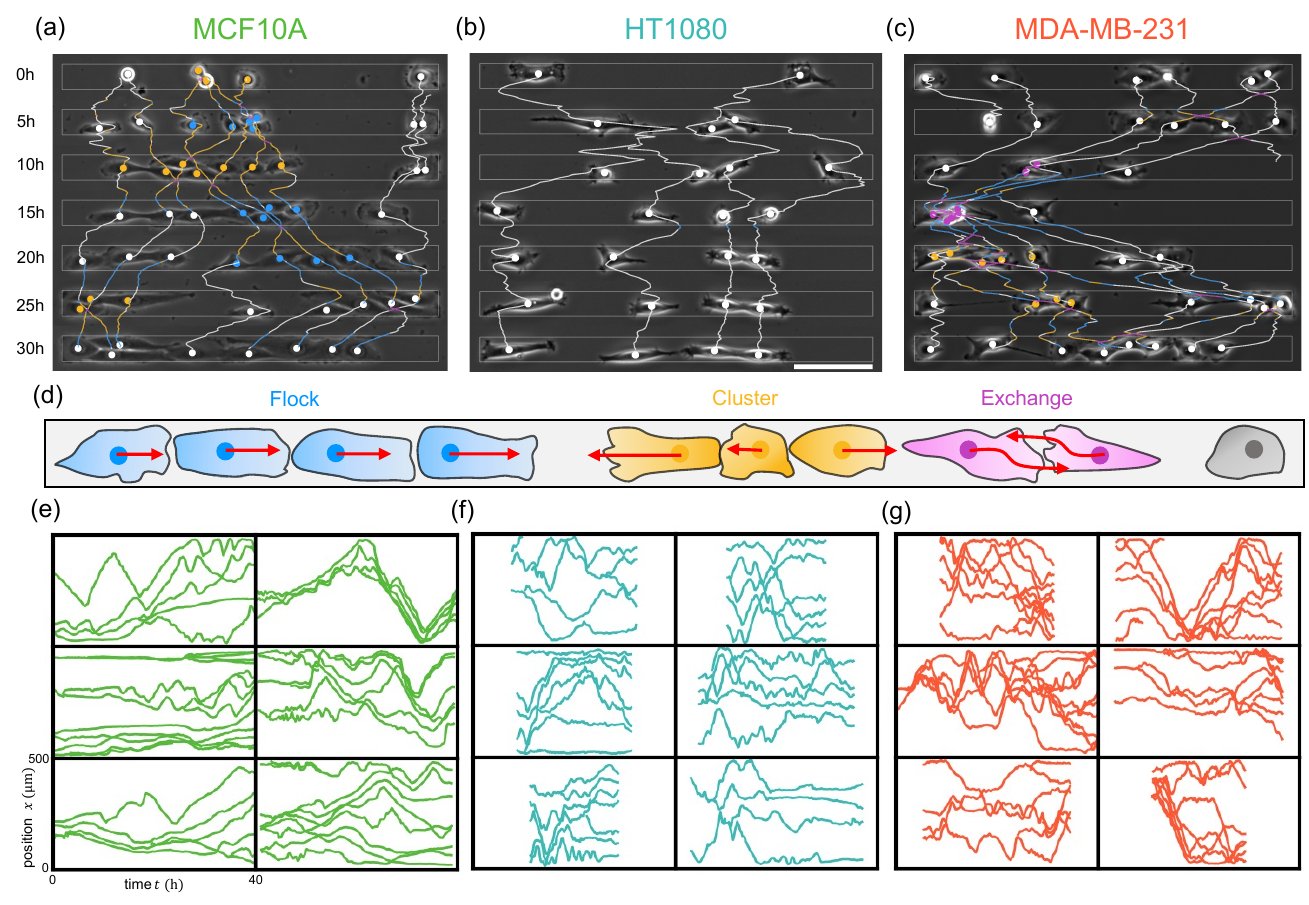}
	\centering
	\caption{
		\textbf{Multicellular behavior across multiple cell types on micropatterns.} \textbf{(a)-(c)} Time-lapse bright-field microscopy images of 1D micropatterned tracks with epithelial MCF10A cells (a), mesenchymal fibrosarcoma HT1080 cells (b) and mesenchymal breast cancer MDA-MB-231 cells (c). Trajectories show the 1D projected cell position over time and therefore do not necessarily pass through the apparent cell center. White lines indicate cell trajectories obtained from cell tracking (Methods~\ref{app:tracking}). Cells in a flock are blue, cells in a cluster without coherent motion are orange, and cells exchanging positions are purple (Methods~\ref{app:methods_behavior_statistics}). Scale bar in (b) is $100\,\si{\micro\meter}$ and applies to all panels (a)-(c). For movies, see Supplementary Section 12. \textbf{(d)} Schematic of collective behaviors observed in our experiments. \textbf{(e)-(g)} Selection of representative examples of experimental cell trajectories. Axis labels in (e) apply to all trajectories.
		}
	\label{Fig1}
\end{figure*}

\subsection*{High-throughput experimental assay for multicellular dynamics}
To learn cell-cell interactions in multicellular systems, we use a high-throughput micropatterned lane assay that minimizes confounding spatial complexity, yet preserves key behaviors of multicellular migration (Methods~\ref{app:Micropatterning},~\ref{app:cellculture}). This environment (Fig.~\ref{Fig1}(a)-(d)) promotes frequent cell contacts and mimics native tissue structures such as aligned collagen fibers or narrow stromal paths \cite{doyle2013dimensions, kraning2013microfabricated}. We study three human cell types that exhibit distinct collective behaviors: non-cancerous epithelial breast cells (MCF10A) typically showing coherent collective motion \cite{brandstatter2023curvature,wong2014collective}, mesenchymal fibrosarcoma cells (HT1080) that tend to avoid sustained contact with neighboring cells \cite{brandstatter2025data}, and mesenchymal breast cancer cells (MDA-MB-231) that have been described as maintaining an individualistic migration style even when part of a group \cite{lee2021distinct}.

We initially seed $2$–$5$ cells of a single cell type on tracks of length $L = 500\ \si{\micro\meter}$ and width $w = 35\ \si{\micro\meter}$. The track width restricts lateral motion, resulting in predominantly quasi-1D dynamics (Supplementary Section 5.1), while still allowing cells to deform and exchange positions. In contrast to studies on transient cell-cell collisions \cite{scarpa2013novel,Desai2013}, we monitor the system over long timescales of up to 48 hours. During this time, cells proliferate to form a system of up to $N_\mathrm{cells}=15$ migrating cells exploring the full track, which we record to obtain microscopy videos of their collective behaviors (Fig.~\ref{Fig1}(a)–(c), Supplementary Section 12). Cell proliferation thus naturally samples a range of cell densities within each experiment, allowing us to investigate how cell density affects multicellular behaviors~\cite{Vedel2013, 10.1098/rsif.2020.0143,angelini2011glass,castro2016clustering,Garcia2015d}. At high cell densities, the mean spacing between cells falls below their typical length $\langle \ell \rangle$, leading to forced lateral stacking. To maintain quasi-1D dynamics, we restrict our analysis to a packing density $\phi = \langle \ell \rangle N_\mathrm{cells} / L \leq 1$ (Supplementary Section 6.1). 

To quantify multicellular dynamics, all cell nucleus positions are tracked over time (Methods~\ref{app:tracking}), reducing complex migration patterns to low-dimensional cell trajectories (Fig.~\ref{Fig1}(e)-(g)). The set of cell trajectories provides an effective representation of whole-cell movements, capturing both individual cell motility---such as speed and directional persistence---and how this motility is altered in the presence of neighboring cells. Repeated experiments yield a dataset of approximately $500 \si{\hour}$ of multicellular trajectories for each cell type, spanning a broad range of cell densities (Supplementary Section 6). This large dataset is key, as it will enable our inference approach to separate dynamics caused by cell-cell interactions from single-cell dynamics and inherent stochasticity.

\subsection*{Cells exhibit distinct multicellular behaviors}
\begin{figure*}[t!]
	\includegraphics[width=0.98\textwidth]{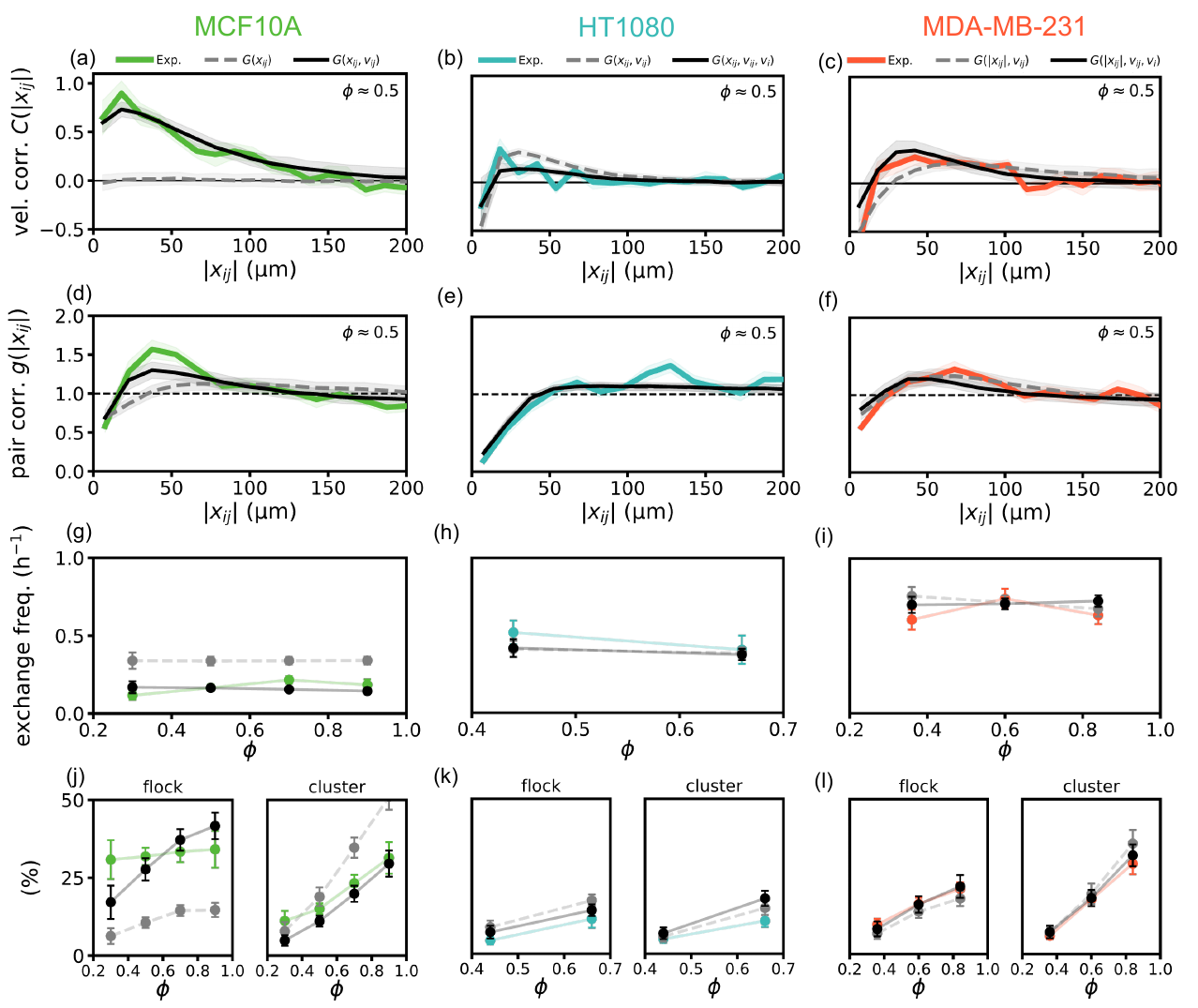}
	\centering
	\caption{
		\textbf{Quantitative analysis of multicellular migration dynamics.} \small \textbf{(a)-(c)} Normalized velocity cross-correlation function $C(|x_{ij}|;\phi)$ for epithelial MCF10A (a), mesenchymal fibrosarcoma HT1080 (b) and mesenchymal breast cancer MDA-MB-231 cells (c) at fixed packing density $\phi\approx 0.5$ (for different densities see Supplementary Fig. S9(a)-(c)). Colored lines show the experimental data; black and gray lines indicate theory results (see legends above panels). \textbf{(d)-(f)} Pair correlation function $g(|x_{ij}|;\phi)$ for the three cell types at density $\phi\approx 0.5$ (for different densities see Supplementary Fig. S9(d)-(f)). \textbf{(g)-(i)} Exchange frequency, defined as the number of exchange events per unit time cells spend in close proximity to each other (Methods~\ref{app:methods_behavior_statistics}), shown as a function of density $\phi$. Error bars indicate the standard error of the mean (s.e.m.) found through bootstrapping (Methods~\ref{app:methods_statistics}). \textbf{(j)-(l)} Percentage of cells that are part of a flock or cluster as a function of the packing density $\phi$, with error bars indicating the s.e.m. found through bootstrapping.
		}
	\label{Fig2}
\end{figure*}

To characterize collective cell behaviors, we start by quantifying their velocity correlations. We compute the normalized velocity cross-correlation as a function of the intercellular distance $|x_{ij}| = |x_i-x_j|$:
\begin{equation}
C(|x_{ij}|; \phi) = \frac{\langle v_i v_j \rangle_{\phi,|x_{ij}|} }{ \langle |v| \rangle_{\phi}^{2}}.
\end{equation}
Here, $\langle ... \rangle_{\phi, |x_{ij}|}$ denotes time and ensemble averages at packing density $\phi$ and distance $|x_{ij}|$ (Methods~\ref{app:methods_behavior_statistics}). For typical densities $\phi \approx 0.5$, we find marked differences between cell types. The epithelial cells show strictly positive velocity correlations that decay over approximately $150 \si{\micro\meter}$ (Fig.~\ref{Fig2}(a))---about 3 cell lengths (Supplementary Section 6.1)---consistent with previous work on epithelial cells \cite{sepulveda2013collective, Garcia2015d}. In contrast, fibrosarcoma and breast cancer cells show negative correlations at short intercellular distances and positive but smaller correlations at longer distances (Fig.~\ref{Fig2}(b),(c)). Notably, these velocity correlations are not noticeably affected by density (Supplementary Fig. S9(a)-(c)), demonstrating that epithelial cells move in a correlated manner, whereas mesenchymal cells display weaker coherence and both positive and negative correlations, highlighting key differences in how these cells organize their relative motion.

Next, we compute the pair correlation function $g(|x_{ij}|; \phi)$ to characterize cellular arrangement at different densities (Methods~\ref{app:methods_behavior_statistics}). Epithelial cells exhibit a peaked pair correlation function, indicating that cells tend to maintain an intercellular distance of $\sim 40\ \si{\micro\meter}$ along with low correlations at short distances, consistent with mutual exclusion (Fig.~\ref{Fig2}(d)). Breast cancer cells are more loosely spaced, with a broader peak around $60\ \si{\micro\meter}$ and similarly low pair correlations at short distances (Fig.~\ref{Fig2}(f)). In contrast, the fibrosarcoma cells exhibit no peak in the pair correlation function and have an even lower value at small intercellular distances (Fig.~\ref{Fig2}(e)), indicating almost no overlap and the absence of a characteristic intercellular distance. For all three cell types, the overall shape of the pair correlation function remains consistent across densities, although the peak height for epithelial cells decreases with $\phi$ (Supplementary Fig. S9(d)-(f)). Taken together, epithelial and breast cancer cells maintain characteristic spacing, while fibrosarcoma cells do not and exhibit no close contact.

To gain further insight into the behaviors underlying these correlation functions, we analyze how cells behave near other cells and how this behavior changes across densities. One characteristic behavior is that cells can exchange positions by sliding past one another (Fig.~\ref{Fig1}(a)-(d)). To quantify this, we define the exchange frequency as the number of position exchange events divided by the time cells spend near each other (Methods~\ref{app:methods_behavior_statistics}), which is largely insensitive to $\phi$ (Fig.~\ref{Fig2}(g)-(i)). Fibrosarcoma and breast cancer cells exchange positions more frequently than epithelial cells, consistent with the negative velocity correlations of cancer cells at short distances.

Finally, we examine multicellular behaviors, classifying cells as individual or grouped, and assessing whether these groups move coherently (flock) or not (cluster) (Methods~\ref{app:methods_behavior_statistics}). As may be expected \cite{theveneau2013collective,jain2020role}, epithelial cells predominantly form flocks and do so more often than mesenchymal cancer cells, which show roughly equal flocking and clustering behavior (Fig.~\ref{Fig2}(j)-(l)). As density increases, epithelial cells cluster more often while flocking remains roughly constant. In contrast, breast cancer cells both flock and cluster more often with increasing density, revealing their ability to move collectively even when densely packed. Fibrosarcoma cells rarely form groups of either kind, even at higher densities.

Taken together, all cell types show short-range mutual exclusion and velocity correlations that extend over multiple cell lengths. However, they differ markedly in their exchange frequency, clustering, and flocking. These contrasting multicellular migration behaviors suggest distinct underlying cell-cell interactions. 

\begin{figure*}[t!]
	\includegraphics[width=\textwidth]{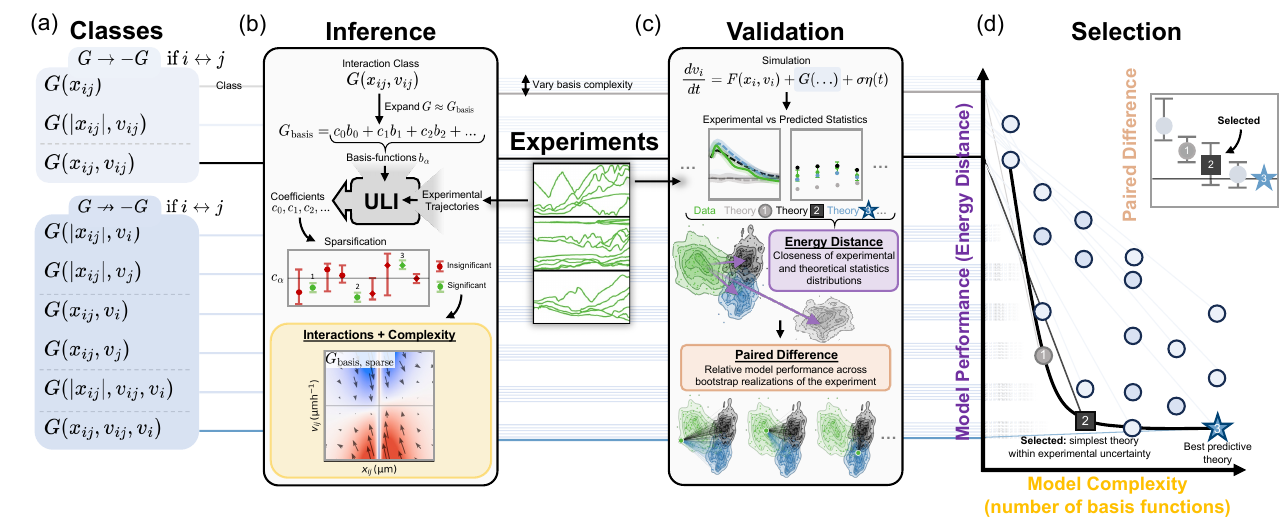}
	\centering
	\caption{
		\textbf{Many-body stochastic inference framework.} \textbf{(a)} A comprehensive list of possible interaction classes that we consider, grouped according to whether they obey action-reaction symmetry. Dashed lines separate classes with the same number of vectorial dependencies. \textbf{(b)} Inference procedure used to learn the interaction terms defined in panel (a). Interactions are expanded in basis functions and functions $g_\alpha$ are inferred using Underdamped Langevin Inference (ULI) from experimental cell trajectories. Basis functions whose coefficients are not resolved by bootstrap uncertainty are removed iteratively. \textbf{(c)} Validation step of our inference procedure. We simulate each inferred equation of motion and compute the collective behavior statistics from Fig.~\ref{Fig2}. Experimental bootstrap realizations and repeated simulations give two distributions of these combined statistics, which we compare using energy distance. To assess uncertainty from finite experimental sampling, we compute the energy distance of two models to the same experimental bootstrap realization and take their difference (paired energy-distance difference $\Delta D_{\mathrm E}$). \textbf{(d)} Model selection based on predictive agreement and interaction complexity, measured by the number of basis functions retained after sparsification. The black line connects models on the complexity front: each has a lower energy distance than every model with fewer retained basis functions. We compare simpler models on this front with the lowest-distance model (blue star) using the paired differences from (c). The black square marks the simplest model that outperforms the lowest-distance model in at least 5\% of experimental bootstrap realizations (inset).
		}
	\label{Fig3}
\end{figure*}

\subsection*{Inferring multicellular dynamics}
Our goal is to learn the cell-cell interactions underlying these multicellular migration behaviors from the data. To this end, we infer directly from measured multicellular trajectories an equation of motion that describes how cell encounters alter cell velocity. Despite operating at low Reynolds numbers, the position of a migrating cell exhibits effectively underdamped dynamics because reorienting or reversing motion requires time to remodel the internal organization of the cell \cite{jain2020role,bruckner2024learning, camley2014polarity}. Thus, following studies on single cell migration \cite{selmeczi2005cell}, pairs of interacting migrating cells \cite{bruckner2021learning,brandstatter2025data}, and multicellular migration \cite{sepulveda2013collective,chepizhko2016bursts}, we posit that cell accelerations can be described by a general underdamped Langevin equation,
\begin{equation}\label{eq:GLE}
    \dot{v}_i = F(x_i, v_i) + M[v_i](t) + \sum_{j\neq i} G(x_i, x_j,v_i,v_j) + \sigma \eta_{i}(t).
\end{equation}  
Cells are modeled as identical particles with three contributions: single-cell behavior, pairwise interactions, and dynamical noise that captures their inherent stochasticity. Single-cell behavior is described by a one-body Markovian term $F(x_i, v_i)$ and a memory term $M[v_i](t)$, whose structure is independently constrained from single-cell trajectories (Methods~\ref{app:methods_basis},\ref{app:methods_memory}). Cell stochasticity is incorporated as Gaussian white noise $\eta_{i}(t)$ with amplitude $\sigma$ \cite{selmeczi2005cell,bruckner2021learning,sepulveda2013collective,chepizhko2016bursts}. Finally, pairwise interactions are captured by a two-body term $G(x_i, x_j, v_i, v_j)$, which we here assume---and later check---to be independent of density in our experiments. To learn dynamical cell-cell interaction rules, we infer $G$ directly from the experimental data. 

Because it is unknown how the unobserved biochemical and mechanical processes coarse-grain to $G$ as a function of cell positions and velocities, we cannot know \textit{a priori} on which dynamic variables $G$ depends, much less what the functional form of these dependencies is. As a starting point, we first constrain possible dependencies by imposing translational symmetry such that interactions depend on $x_{ij} = x_i-x_j$ rather than absolute positions. We further require $G$ to depend on $|x_{ij}|$, allowing interactions to vanish as $|x_{ij}|\rightarrow\infty$. Within these constraints, we propose possible \textit{interaction classes} with an increasing number of directional dependencies, allowing cells to exhibit more complex interactive responses (Fig.~\ref{Fig3}(a)). The simplest interaction classes consist of $G(x_{ij})$, $G(|x_{ij}|, v_{ij})$, and $G(x_{ij}, v_{ij})$, where $v_{ij}$ is the relative velocity $v_{ij} = v_i - v_j$. These classes are reciprocal, enforcing action-reaction symmetry ($G \rightarrow -G$ under $i \leftrightarrow j$). However, owing to their active nature, cells may exhibit nonreciprocal interactions depending explicitly on individual velocities $v_i$ or $v_j$. Therefore, we also consider interaction classes such as $G(|x_{ij}|, v_{j})$, $G(|x_{ij}|, v_{ij}, v_{i})$, and the most general class $G(x_{ij}, v_{ij}, v_i)$, which is equivalent to $G(x_{ij}, v_{ij}, v_{j})$ and $G(x_{ij}, v_{i}, v_{j})$ (Supplementary Section 7). We choose to focus on $G(x_{ij}, v_{ij}, v_i)$ as its $v_{ij}$-dependence enables direct comparison to reciprocal models, while its $v_i$-dependence encodes the assumption that a cell may have more ready access to its own velocity than to that of its neighbors. This procedure leads to a comprehensive range of interaction classes (Fig.~\ref{Fig3}(a)). 

For each of these interaction classes, the functional form of their dependencies is still unknown. We therefore infer the functional form for each class directly from the trajectories using Underdamped Langevin Inference (ULI) (Fig.~\ref{Fig3}(b)) \cite{bruckner2020inferring}. ULI uses stochastic estimators to rigorously infer the Markovian components of Langevin equations from positions, velocities, and accelerations contained in the measured cell trajectories, while correcting for biases from discrete time sampling and measurement error (Methods~\ref{app:methods_inference}). We extend these estimators to correct for the contribution of the memory term (Supplementary Section 2).

A key step in Langevin inference is to expand the interaction models into basis functions (Supplementary Section 1.1.2). We choose basis functions that respect the symmetries and asymptotic behaviors of the interactions, and we separate their spatial and directional dependence for interpretability. For example, for the most general interaction class we write 
\begin{equation}\label{eq_expansion}
    G(x_{ij},v_{ij},v_i) = \sum_{\alpha}g_\alpha(|x_{ij}|)P_n^\alpha(x_{ij},v_{ij},v_i),
\end{equation}
where $g_\alpha(|x_{ij}|)$ are kernels that describe the spatial structure of the interaction. To capture a broad range of interactions, we choose to expand $g_\alpha(|x_{ij}|)=\sum_\beta c_{\alpha\beta}e^{-|x_{ij}|/r_\beta}$ into exponentials with varying decay lengths, allowing complex spatial structure while imposing a finite interaction range. Furthermore, $P_n^\alpha(x_{ij}, v_{ij}, v_i)$ denotes the $\alpha$th term of an $n$th-order polynomial in the three vectorial dependencies, which define the direction in which the kernels act. Reflection symmetry is enforced by retaining only odd terms in this polynomial. Importantly, other flexible basis expansions yield equivalent kernels, demonstrating that our results are not sensitive to the particular choice of basis (Supplementary Section 9.3). The coefficients $c_{\alpha\beta}$ are inferred from the experimental trajectories through ULI, yielding the functions $g_\alpha(|x_{ij}|)$ for each interaction class (Fig.~\ref{Fig3}(b)). To avoid overfitting, we iteratively remove basis functions whose coefficients are not resolved due to bootstrap uncertainty (Methods~\ref{app:methods_inference}). The final number of retained basis functions $N_b$ quantifies the complexity of the resulting model. In summary, using only short-timescale dynamic information from trajectories, we infer the full form of both deterministic and stochastic parts of the multicellular equation of motion for a specific interaction class.  

\subsection*{Model Validation and Selection}
Following our inference procedure, we use short-time statistics of cell motion to learn---in parallel---many different models with varying complexity from the same dataset. However, the resulting models do not necessarily reproduce the long-time or large-scale multicellular behaviors observed in our experiments, requiring a model validation step. Because models differ in interaction class (Fig.~\ref{Fig3}(a)) and complexity, they can predict distinct multicellular behaviors. These behaviors were not used to infer the models, yet they are precisely what a model of cell--cell interactions should capture, making them a natural basis for model selection. In the final step of our learning procedure, we therefore use the collective behavior statistics observed in our simulations and experiments (Supplementary Section 4) to select a minimal yet predictive model of cell–cell interactions.

To perform model selection, we note that the inferred models and the experiments are characterized by inherently variable behavior (variance in Fig.~\ref{Fig2}). To determine which model best captures both the average behavior and the observed variability, we first use bootstrapping to quantify the experimental variability of multicellular migration: We randomly resample whole experiments with replacement and recompute the statistics of Fig.~\ref{Fig2} for each bootstrap realization (Fig.~\ref{Fig3}(c), Methods~\ref{app:methods_statistics}). Next, we combine the statistics of each bootstrap realization into an observable vector $s_{\mathrm e}$, whose elements correspond to the values of the discretized experimental correlation functions and multicellular behavior statistics shown in Fig.~\ref{Fig2}. Importantly, each element is normalized by its bootstrap standard deviation and by the square root of the number of discrete bins in that statistic, such that each statistic is weighted equally. The bootstrapped ensemble of these vectors characterizes how inherent variability in cell behaviors affects our multicellular behavior statistics.

The ensemble of statistics vectors $s_{\mathrm m}$ for a model can be readily obtained by repeatedly simulating multicellular trajectories and computing all the resulting statistics for different realizations (Methods~\ref{app:methods_statistics}). To score how well the model predicts the experimental statistics distribution, we use the so-called energy distance~\cite{rizzo2016energy}, defined as
\begin{equation}\label{eq_energy_distance}
D_{\mathrm E}({s_{\mathrm e}},{s_{\mathrm m}})
= 2\overline{d(s_{\mathrm e},s_{\mathrm m})}
- \overline{d(s_{\mathrm e},s_{\mathrm e}')}
- \overline{d(s_{\mathrm m},s_{\mathrm m}')}.
\end{equation}
Here, $s_{\mathrm e}, s_{\mathrm e}'$ and $s_{\mathrm m}, s_{\mathrm m}'$ denote independent draws from each behavior statistics cloud, the overline denotes an average over all pairs, and $d(\cdot,\cdot)$ is the Euclidean distance between two statistic vectors (Methods~\ref{app:methods_model_selection}, Supplementary Section 4.2). The first term averages all experimental-model distances, the second term subtracts the internal spread of the experimental cloud, and the third term subtracts the internal spread of the model cloud. The energy distance is zero for identical distributions and grows as their means or spreads differ. Thus, the model with the lowest energy distance is the closest predictive match to the experimental distribution.

However, the model with the lowest energy distance is not necessarily the best choice. Because energy distances are estimated from finite data, the best model's advantage over other models may not persist in a repeated experiment, in which a model with fewer basis functions might provide the closest predictive match. To identify such models, we construct a complexity front consisting of all models for which there exists no simpler model with lower energy distance (Fig.~\ref{Fig3}(d)). For each front model $m$, we ask whether the lowest-energy distance model $m^*$ would still outperform it if the experiment were repeated. Since the experiment cannot be repeated indefinitely, we treat each experimental bootstrap realization $s_{\mathrm e}^{(k)}$ as a plausible repeat and compute the energy distance of both models to the same realization, which yields the paired energy-distance difference~\cite{gneiting2007proper}
\begin{equation}\label{eq_score_difference}
\Delta D_{\mathrm E}^{(k)} = D_{\mathrm E}\big(s_{\mathrm e}^{(k)},s_{\mathrm m}\big) - D_{\mathrm E}\big(s_{\mathrm e}^{(k)},s_{\mathrm m^*}\big).
\end{equation}
Here, $D_{\mathrm E}\big(s_{\mathrm e}^{(k)},s_{\mathrm m}\big)$ is the energy distance between a single bootstrap realization $k$ and the predictions of model $m$ (Methods~ \ref{app:methods_model_selection}). This paired difference $\Delta D_{\mathrm E}^{(k)}$ is thus negative when $m$ predicts realization $k$ better than $m^*$. Averaged over realizations, $\Delta D_{\mathrm E}^{(k)}$ recovers the energy-distance gap between $m$ and $m^*$, while its spread shows how likely their ranking is to change in a repeated experiment. We consider $m$ not significantly worse than $m^*$ if at least 5\% of these differences are negative, which corresponds to a one-sided bootstrap test with $p = 0.05$ (Methods~ \ref{app:methods_model_selection}). Among all front models that pass this test, we select the one with the fewest basis functions (black square in Fig.~\ref{Fig3}(d)), yielding the minimal model among all well-performing candidates.

\begin{figure*}[t!]
	\includegraphics[width=\textwidth]{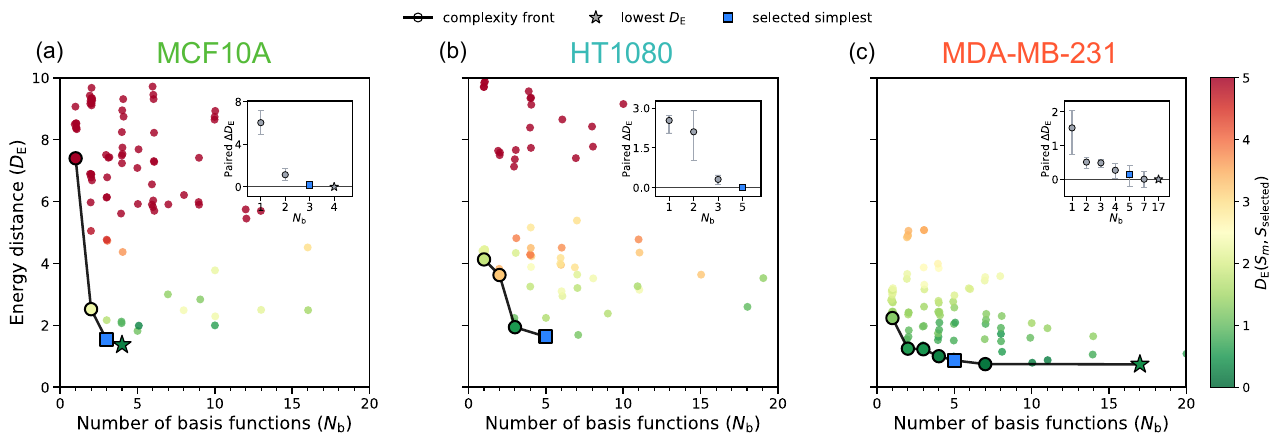}
	\centering
	\caption{
\textbf{Model selection from collective behavior statistics.}
\textbf{(a)--(c)} Energy distance $D_{\mathrm E}$ between the experimental behavior-statistic distribution and each inferred model for MCF10A (a), HT1080 (b), and MDA-MB-231 (c). Each point represents one inferred interaction model, with the horizontal position indicating the number of retained non-zero basis functions $N_b$ and the vertical position indicating its energy distance to the experimental bootstrap cloud. Point colors indicate the energy distance between the model statistics and those of the selected simplest model, $D_{\mathrm E}(s_{\mathrm m},s_{\mathrm{selected}})$. The black line connects the complexity front, consisting of models that improve on all models with fewer retained basis functions. The star marks the model with the lowest energy distance to the experimental statistics, while the blue square marks the selected simplest model. Insets show the paired energy-distance difference $\Delta D_{\mathrm E}$ between each front model and the lowest-distance model, evaluated for the same experimental bootstrap realizations. Points indicate the median paired difference and error bars show the 5th--95th percentile range. A negative value indicates that the simpler model predicts that experimental realization more accurately. A simpler model is selected when at least 5\% of the bootstrap realizations favor it. Note that in panels (a) and (b), some models have $D_{\mathrm E} > 10$ and are off-axis.
}
	\label{fig:3.5}
\end{figure*}

\begin{figure*}[t!]
	\includegraphics[width=\textwidth]{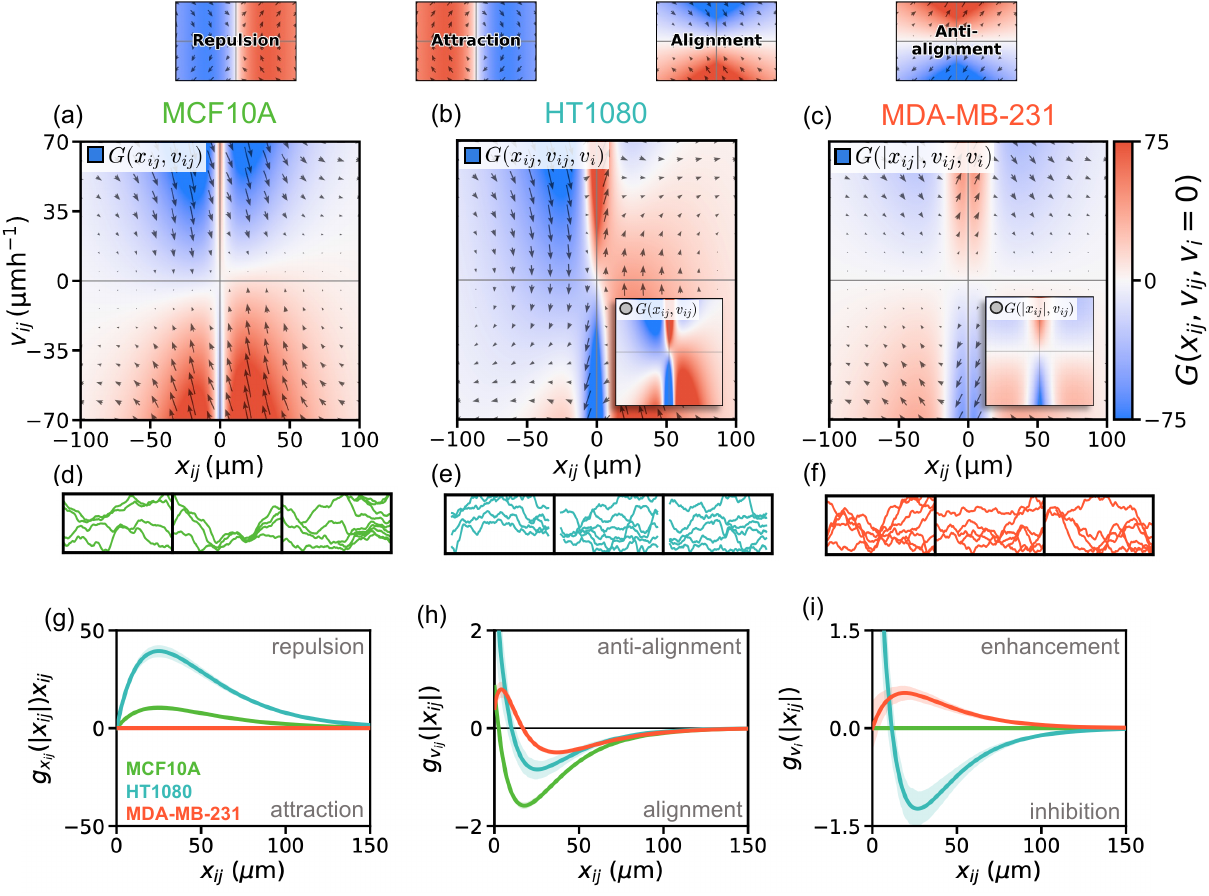}
	\centering
	\caption{
		\textbf{Inferred interaction rules underlying multicellular dynamics.} \textbf{(a)-(c)} Reciprocal part ($g_{x_{ij}}(|x_{ij}|) x_{ij} +g_{v_{ij}}(|x_{ij}|) v_{ij}$) of the selected interaction model for each cell type. Color indicates the interaction-induced acceleration $G$ of cell $i$, and the arrows depict the time evolution of the relative position and velocity, with vectors $\left[v_{ij}\, , \, G(x_{ij},v_{ij}) - G(-x_{ij},-v_{ij})\right]^\top$. Schematic heat maps at the top indicate how pure repulsion, attraction, alignment, and anti-alignment would look in this representation. The insets in (b) and (c) show inferred dynamics from the purely reciprocal interaction classes (Supplementary Table S4), which are similar to the reciprocal parts of the full models. \textbf{(d)-(f)} Simulated trajectories for the selected interaction models for the three cell types. \textbf{(g)-(i)} Inferred kernels for all cell types as a function of relative position. (g),(h) show the reciprocal kernels $g_{x_{ij}}(|x_{ij}|) x_{ij}$ and $ g_{v_{ij}}(|x_{ij}|)$, and the nonreciprocal interaction kernel $g_{v_{i}}(|x_{ij}|)$ is shown in (i). Shaded area indicates the 68\% bootstrap percentile confidence interval, approximately equal to 1 s.e.m.
		}
	\label{Fig4}
\end{figure*}

\subsection*{Inferred cell-cell interactions capture multicellular dynamics}\label{sec:inference_results}
Having developed our many-body statistical inference pipeline and verified its self-consistency (Supplementary Section 9.1), we now determine the interaction rules that drive multicellular migration in our experiments. Applying our model selection procedure, we find that for all three cell types, along the complexity front, the energy distance drops steeply over the first few retained basis functions and then plateaus, indicating that a small number of basis functions suffices to capture the experimental statistics (Fig.~\ref{fig:3.5}(a)-(c)). For HT1080 cells, we select the lowest-distance model (Fig.~\ref{fig:3.5}(b)), whereas for MCF10A and MDA-MB-231 cells, simpler models on the front perform comparably to the lowest-distance models (insets in Fig.~\ref{fig:3.5}(a),(c)) and are therefore selected instead. We note that the lowest-distance and simpler models on the front predict very similar multicellular behaviors (colors of the dots in Fig.~\ref{fig:3.5}(a),(c)), indicating that the increased complexity improves the same behavioral predictions, rather than producing distinct behaviors. 

All selected models are first order in the directional dependencies, which allows us to write them in an intuitive form
\begin{equation}\label{eq:nonrecip:_decomp}
G(x_{ij}, v_{ij}, v_i) = g_{x_{ij}}(|x_{ij}|) x_{ij} + g_{v_{ij}}(|x_{ij}|) v_{ij}+ g_{v_{i}}(|x_{ij}|) v_{i}.
\end{equation}
Here, $g_{x_{ij}}(|x_{ij}|)$ acts along the relative position $x_{ij}$, with positive values corresponding to repulsion and negative values to attraction. Similarly, $g_{v_{ij}}(|x_{ij}|)$ acts along the relative velocity $v_{ij}$, meaning negative values correspond to velocity alignment and positive values to anti-alignment. In the nonreciprocal contribution, the kernel $g_{v_{i}}(|x_{ij}|)$ acts along a cell's own velocity $v_{i}$, meaning that positive values correspond to an acceleration (\textit{velocity enhancement}) and negative values to a deceleration (\textit{velocity inhibition}), allowing a cell to adjust its motion upon contact with another cell based solely on its own velocity $v_{i}$.

The selected interaction classes differ qualitatively between cell types: the model for epithelial cells is purely reciprocal, with $G(x_{ij}, v_{ij})$ including repulsion and velocity alignment (Fig.~\ref{Fig4}(a),(g),(h)). We confirm our earlier assumption that these interactions remain similar across density (Supplementary Fig. S10(a)) and capture the experimental statistics across densities, including large positive velocity correlations, the peak in the pair correlation, low exchange frequency, and flock and cluster percentages (Fig.~\ref{Fig2}(a),(d),(g),(j)). In contrast, a simpler model without dependence on relative velocity $v_{ij}$ fails to describe the observed behavior (Fig.~\ref{Fig2}(a),(d),(g),(j)), showing that velocity alignment is essential to capture multicellular behaviors of epithelial cells, in line with earlier results obtained from two-cell collision behaviors \cite{bruckner2021learning}.

For fibrosarcoma cells, we infer strong repulsion compared to epithelial cells, while for breast cancer cells we infer no repulsion/attraction ($g_{x_{ij}}=0$) interactions (Fig.~\ref{Fig4}(b),(c),(g)). However, both cell types exhibit a distinctive combination of short-range velocity anti-alignment and long-range alignment interactions (Fig.~\ref{Fig4}(b),(c),(h)). These opposing alignment interactions can qualitatively explain the fibrosarcoma's and breast cancer cell's seemingly conflicting behavior of both negative and positive velocity correlations (Fig.~\ref{Fig2}(b),(c)) as well as their ability to flock despite frequent cell exchanges (Fig.~\ref{Fig2}(h),(i),(k),(l)). 

Remarkably, both mesenchymal cell types exhibit nonreciprocal interactions, with interaction class $G(x_{ij}, v_{ij}, v_i)$ for fibrosarcoma cells and $G(|x_{ij}|, v_{ij}, v_{i})$ for breast cancer cells, which depend explicitly on the velocity of the cell. The kernel $g_{v_i}$ reveals that fibrosarcoma cells exhibit velocity inhibition, whereas breast cancer cells exhibit velocity enhancement (Fig.~\ref{Fig4}(i)). These inferred interactions are again consistent across densities (Supplementary Fig. S10(b),(c)), and we observe that velocity enhancement does not lead to unphysical exponential velocity growth (Supplementary Section 6.4). Together with the reciprocal interactions, the nonreciprocal interactions yield a good prediction of the velocity correlations and flock percentages for both cell types across densities (black lines in Fig.~\ref{Fig2}(b),(c),(k),(l), Supplementary Fig. S10(b),(c)). These results suggest that nonreciprocal interactions are important for capturing the collective dynamics of both mesenchymal cell types, raising the question of which behavioral features they explain beyond those captured by purely reciprocal models.

To gain initial insight into the contribution of these nonreciprocal interactions, we compare the selected mesenchymal models with the best purely reciprocal model for each cell type:  $G(x_{ij}, v_{ij})$ for fibrosarcoma and $G(|x_{ij}|, v_{ij})$ for breast cancer cells (gray dashed lines in Fig.~\ref{Fig2}(b),(c),(h),(i),(k),(l); insets in Fig.~\ref{Fig4}(b),(c)). The reciprocal terms of these models are again selected using our model selection framework and interestingly retain the same qualitative structure as those of the selected nonreciprocal models, with short-range anti-alignment and long-range alignment (Fig.~\ref{Fig4}(b),(c) and insets). However, without the nonreciprocal component, the positive velocity correlations and flock percentages are overestimated for fibrosarcoma cells and underestimated for breast cancer cells (Fig.~\ref{Fig2}(b),(c),(k),(l)). These features are present across densities (Supplementary Fig. S17(b),(c)) and match the sign of the inferred interaction: velocity inhibition lowers the degree of coherent motion in fibrosarcoma cells, while velocity enhancement raises it in breast cancer cells. These comparisons indicate that nonreciprocal interactions regulate collective coordination, but do not explain how their interplay with alignment and repulsion shapes flock formation and motion.

\begin{figure*}[t!]
	\includegraphics[width=\textwidth]{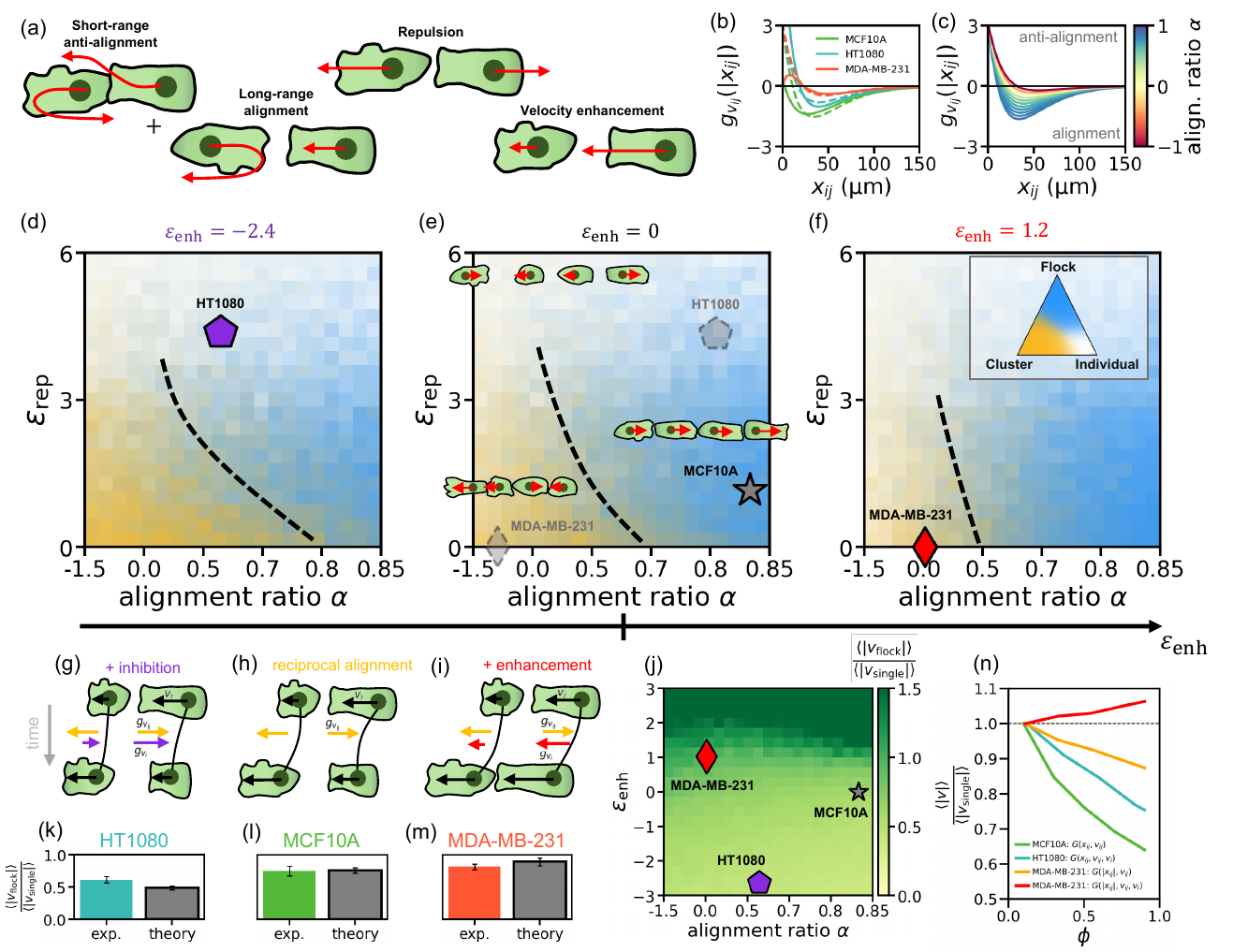}
	\centering
	\caption{
		\textbf{Interaction-collective behavior space and minimal interaction model.} \textbf{(a)} Schematic depiction of the behaviors expected from the four inferred cell-cell interactions. \textbf{(b)} The anti-alignment/alignment kernel $g_{v_{ij}}$ of the three cell types inferred from the experiments (solid lines). We simultaneously fit $g_{v_{ij}}$ of all cell types with the expression $g_{v_{ij}}(|x_{ij}|)=(\varepsilon_\mathrm{align} |x_{ij}| + d)\,e^{-|x_{ij}|/r_\mathrm{align}}$ with $d$ and $r_\mathrm{align}$ constant across cell types (Supplementary Section 11.1). Fit results are shown as dashed lines. \textbf{(c)} The anti-alignment/alignment kernel $g_{v_{ij}}$ used in our minimal model while varying $\varepsilon_\mathrm{align}$. For each value of $\varepsilon_\mathrm{align}$ we compute the alignment ratio $\alpha$ defined as $\alpha = 1-S_+/|S_-|$, where $S_\pm$ is the integral of $g_{v_{ij}}$ above or below zero.  
        \textbf{(d)-(f)} Collective behavior as a function of the alignment ratio $\alpha$ and the repulsion strength $\varepsilon_\mathrm{rep}$ for different values of $\varepsilon_\mathrm{enh}$. Throughout, we indicate a flocking transition by black dashed lines. Symbols indicate the inferred interactions of cell types. In panel (e), we indicate the hypothetical position of fibrosarcoma and breast cancer cells in the purely reciprocal plane. Color map in panel (f) applies to all panels (d)-(f). \textbf{(g)}-\textbf{(i)} Schematic depictions of the effect of velocity inhibition/enhancement interactions on the interaction behavior of cells. Throughout, black arrows indicate the velocity of cells, while colored arrows indicate cell acceleration due to interactions. Upper row indicates initial positions and velocities; lower row indicates final positions and velocities. \textbf{(j)} Flock speed relative to single cell speed as a function of the nonreciprocal interaction strength $\varepsilon_\mathrm{enh}$ and the alignment ratio $\alpha$. Symbols indicate inferred interaction parameters for the three cell types. \textbf{(k)}-\textbf{(m)} Quantitative comparison between flock speeds from experiment (colored) and predicted from the inferred models (black). Error bars indicate the s.e.m. \textbf{(n)} Mean cell speeds of simulations normalized by the mean single cell speed as a function of density.
		}
	\label{Fig5}
\end{figure*}

\subsection*{Nonreciprocal interactions tune the onset and speed of flocks} 
To disentangle the various reciprocal and nonreciprocal interactions and gain conceptual insight into the role of each, we perform a dimensional reduction of the full models to a low-dimensional theory with just three parameters (Supplementary Section 11.1): repulsion strength ($\varepsilon_{\mathrm{rep}}$), trade-off between short-range velocity anti-alignment and long-range alignment (parameterized by $\alpha$), and the strength of velocity inhibition or enhancement ($\varepsilon_{\mathrm{enh}}$) (Fig.~\ref{Fig5}(a)-(c)). Using this reduced model, we link individual interactions to collective behavior: whether cells form flocks, clusters, or move individually. We refer to this mapping as the interaction–collective behavior space (ICBS), in which we indicate all cell types according to their inferred interactions (Fig.~\ref{Fig5}(d)–(f)) and which is robust to the choice of single-cell parameters (Supplementary Section 11.3). At these locations, the predicted collective behaviors match experimental observations (Supplementary Section 11.2), providing credence to this low-dimensional representation.

Using this reduced theory and ICBS, we explore how velocity inhibition and enhancement shape collective behavior. Without nonreciprocal interactions ($\varepsilon_{\mathrm{enh}} = 0$), increasing the velocity alignment ratio $\alpha$ leads to a transition from clusters to flocks (dashed line in Fig.~\ref{Fig5}(e)). Increasing repulsion $\varepsilon_{\mathrm{rep}}$ shifts this transition to lower velocity alignment ratios, likely because it increases intercellular distances where long-range alignment dominates (Fig.~\ref{Fig5}(b),(c)). At very high repulsion, clusters fail to form altogether (Fig.~\ref{Fig5}(e)). Introducing nonreciprocal interactions shifts the transition from clusters to flocks: velocity inhibition ($\varepsilon_{\mathrm{enh}} < 0$) requires stronger velocity alignment (higher $\alpha$), while velocity enhancement ($\varepsilon_{\mathrm{enh}} > 0$) lowers this threshold (Fig.~\ref{Fig5}(d),(f)), clarifying why nonreciprocal models match the experimental flock percentages better (Fig.~\ref{Fig2}(k),(l)). Thus, repulsion and velocity alignment primarily determine the overall mode of collective motion, but velocity inhibition and enhancement tune the onset of flocks by modulating the required alignment strength.

We next ask whether these nonreciprocal interactions also affect flock dynamics. As one would intuitively expect, the reduced model shows that velocity inhibition ($\varepsilon_{\mathrm{enh}} < 0$) decreases flock speed $v_\mathrm{flock}$ and velocity enhancement ($\varepsilon_{\mathrm{enh}} > 0$) increases it, where $v_\mathrm{flock}$ can even exceed the average speed of isolated cells (Fig.~\ref{Fig5}(g)-(j)). In line with this, the measured flock speed of each cell type reflects its nonreciprocal interaction: fibrosarcoma cells with velocity inhibition form slow flocks, epithelial cells without enhancement/inhibition interactions have intermediate flock speed, and breast cancer cells with velocity enhancement form fast flocks (Fig.~\ref{Fig5}(k)–(m)). These results show that nonreciprocal interactions tune the speed of flocks.

Finally, these nonreciprocal interactions affect how dense multicellular systems behave (Supplementary Section 8). Breast cancer cells with velocity enhancement exhibit higher speeds at higher density (Fig.~\ref{Fig5}(n)), consistent with experiments on these cell types \cite{jayatilaka2017synergistic}. In contrast, cell speed in the corresponding reciprocal model and models with mostly repulsion and velocity alignment predict a slowdown (Fig.~\ref{Fig5}(n)), as observed experimentally for epithelial cells \cite{angelini2011glass,castro2016clustering}. Together, these results show that nonreciprocal velocity inhibition and enhancement not only tune the onset and speed of flocks, they also impact average cell speed in multicellular systems.

\subsection*{Discussion}
Cell--cell interactions, such as contact inhibition of locomotion first described over 70 years ago \cite{abercrombie1954observations}, have remained challenging to quantify, particularly in multicellular settings where contributions from multiple neighbors are entangled with intrinsic motility and noise. Many models of collective migration therefore still rely on qualitative descriptions of how cells alter their motion in response to their neighbors \cite{camley2014polarity,d2017contact,fujimori2019tissue,alert2020physical}. With our many-body stochastic inference framework, we move beyond such descriptions and learn---directly from data---quantitative, dynamical cell-cell interactions across distinct cell types. With this approach, we recovered previously reported interactions and discovered new ones.

For epithelial MCF10A cells, the inferred repulsion and velocity alignment interactions reproduce coherent multicellular migration, consistent with observations in cell trains and sheets \cite{bertrand2024clustering,sepulveda2013collective,chepizhko2016bursts, Ron2023a} while also predicting previously reported collective slowing at high densities \cite{angelini2011glass,castro2016clustering,Garcia2015d}. Although inferred from multicellular data, these interactions agree with those inferred from isolated pairs of cells \cite{bruckner2021learning}, indicating that epithelial two-body interactions are conserved in our multicellular setting~\cite{xsg5-cb6l}. In contrast, mesenchymal breast cancer cells were previously described as moving individually in fluid-like assemblies \cite{wong2014collective,castro2016clustering,kang2021novel}. However, we show that these cells exhibit intricate collective coordination through a combination of short-range anti-alignment with long-range alignment. Remarkably, we also uncover nonreciprocal interactions---velocity inhibition and enhancement---that were not resolved by inference on isolated cell pairs \cite{bruckner2021learning,brandstatter2025data}. Such nonreciprocal interactions are a hallmark of active matter, where interactions need not obey action--reaction symmetry \cite{marchetti2013hydrodynamics,Klapp2023Non-reciprocalMatter,bowick2022symmetry, shi2026hamiltonian, Tan2021}. Here, our models show that these interactions can tune the onset and speed of flocks. Furthermore, velocity enhancement leads to breast cancer cells moving faster at higher densities, which may prevent a jammed state even at densities beyond those in our experiment (Supplementary Section 8). This prediction is consistent with experimental observations \cite{lee2021neighbor,jayatilaka2017synergistic}, suggesting that nonreciprocal interactions may provide a dynamical understanding for contact-enhanced cancer cell invasion \cite{kang2021novel, d2017contact}.

Inferring cell--cell interactions from multicellular data bridges the gap between the dynamics of isolated cell pairs and many interacting cells. Cell pairs are mostly studied on confining micropatterns that constrain the range of behaviors interacting cells can exhibit \cite{bruckner2021learning,brandstatter2025data,scarpa2013novel,Desai2013,camley2014polarity}. In contrast, cells on multicellular lanes are less confined and have multiple neighbors, each in possibly different migratory states~\cite{bertrand2024clustering}. This allows interacting cells to exhibit a much broader range of dynamical behaviors. This richer set of dynamical configurations more closely resembles \textit{in vivo} conditions of collective cell migration and may be crucial for resolving interaction terms that remain hidden in pair experiments~\cite{bruckner2021learning}, such as those we identify in mesenchymal cells. In addition, proliferation exposes cells to densities that isolated cell pairs cannot sample, allowing us to ask whether cells adapt their interactions to the number of neighbors or average contact time~\cite{Vedel2013, 10.1098/rsif.2020.0143}. We find that in our experiments they largely do not: the inferred interactions are largely conserved across density, so the density dependence of clustering, flocking, and cell speed emerges from the many-body dynamics rather than from cells adapting their interactions.

Taken together, learning interaction laws from multicellular data rather than imposing them allows our framework to integrate disparate observations of cell behavior into a single quantitative setting. Previous inference approaches have provided important insights through the statistical inference of effective interactions from collective correlations \cite{bialek2012statistical,Agliari2020}, the discovery of hydrodynamic equations governing large-scale collective dynamics \cite{Joshi2022Data-DrivenHydrodynamics,Romeo2021,Supekar2023LearningExperiments,brunton2016discovering}, and neural-network representations of collective cell behavior and interactions \cite{lachance2022learning, yang2024learning, Boyle2026}. What distinguishes our approach is its focus on inferring explicit, physically interpretable stochastic interaction laws at the level of individual cells. Despite substantial advances in dynamical inference \cite{brunton2016discovering,Messenger2022LearningPopulation,Huang2022,Ferretti2020BuildingDatasets,Gerardos2025PASTIS}, extracting such laws from noisy multicellular experiments remains challenging. Our framework overcomes this challenge by learning interaction laws directly from experimental trajectories and selecting models based on their ability to reproduce collective behavior statistics. This approach can be extended beyond the simple lane geometry to two~\cite{sepulveda2013collective,chepizhko2016bursts,castro2016clustering, yang2024learning} or three~\cite{brandstatter2023curvature, cetera2014epithelial, kang2021novel} dimensions allowing us to learn the interactions underlying collective migration in larger and potentially physiologically relevant cell tissues. More broadly, the framework can be applied to other living systems---from cell assemblies to animal groups---offering a route from descriptive to quantitative dynamical understanding.

\subsection*{Acknowledgments} 
We thank D.B. Brückner and P. Ronceray for helpful discussions. Part of this work is funded by the Deutsche Forschungsgemeinschaft (DFG, German Research Foundation)—Project-ID 201269156-SFB 1032 (Project B01 and B12). \\
\\B.J.H. and T.B. contributed equally to this work. E.B., T.B., 
and J.O.R. designed experiments; E.B. performed
experiments; B.J.H. and T.B. analyzed data; B.J.H. and T.B. developed the inference framework, performed simulations, and
statistical inference; B.J.H., T.B. and C.P.B.
interpreted the results and wrote the manuscript. J.O.R. and
C.P.B. conceived this study and led the research. All authors
edited and approved the manuscript.

\subsection*{Competing interests}
The authors declare no competing interests.

\appendix
\renewcommand{\appendixname}{Methods}
\begin{table*}[t!]
 \centering
 \begin{tabular}{|l|l|l|}
  \hline
  \textbf{Cell type} & \textbf{Medium} & \textbf{Origin}\\
  \hline
  MDA-MB-231 H2B-mCherry & L-15 (Gibco) + 10\% FBS (Gibco) & Gift from Betz Lab\\
  \hline
  HT1080 & MEM (Gibco) + 10\% FBS (Gibco) + 5\% CO$_2$ & Authenticated by CLS\\
  \hline
  MCF10A & \parbox[t]{0.5\textwidth}{\raggedright DMEM/F-12 (Gibco) + 5\% horse serum (Gibco), 20~ng/ml human epidermal growth factor (Sigma), 100~ng/ml cholera toxin (Sigma), 10~$\mu$g/ml insulin (Sigma) and 500~ng/ml hydrocortisone (Sigma) + 5\% CO$_2$\strut} & ATCC\\
  \hline
 \end{tabular}
 \caption{Cell culture conditions of the different cell lines and their origin.}
 \label{cellcultur}
\end{table*}
\section*{Methods}
\subsection{Micropatterning and sample preparation}
\label{app:Micropatterning}
We employ a micropattern structure with a track-like design with a width of $35\, \si{\micro\meter}$ and a length of $500\, \si{\micro\meter}$. For micropatterning, we employ a photopatterning technique using the PRIMO module (Alvéole). For background passivation of the µ-dish (ibidi), a drop of $0.01\%$ (w/v) PLL (Sigma-Aldrich) is added and incubated for 30 min. Afterward, the sample is rinsed with HEPES buffer (Sigma-Aldrich) and 100 mg/ml PEG-SVA (LaysanBio) is incubated for $1\, \si{\hour}$ at room temperature. The passivated dish is photopatterned by employing the PRIMO module mounted on an automated inverted microscope (Nikon Ti Eclipse). The photoinitiator PLPP (Alvéole) is added to the dish. The track pattern was placed on top of the dish via the Leonardo software (Alvéole) and illuminated with UV light with a dose of $15\, \si{\milli\joule}/\si{\milli\meter}^2$. The dish is washed with milliQ water and rehydrated with PBS for 5 min followed by an incubation with 20 µg/ml of labeled Fibronectin-Alexa647 (YO-protein, ThermoFisher) for 15 minutes at room temperature. 

\subsection{Cell culture}\label{app:cellculture}
The individual culture conditions are shown in Table~\ref{cellcultur}. All cells are grown at $37^\circ\text{C}$ in an atmosphere containing $5\%$ $\mathrm{CO_2}$. Before passaging, cells are washed and treated with accutase for $5  \, \si{\min}$. The cell solution is centrifuged at $800$ r.c.f. for $3 \, \si{\min}$ and the cells are resuspended in medium. Approximately $10,000$ cells are added into the micropatterned µ-dish and left to adhere for up to $4\, \si{\hour}$ in the incubator. After this incubation period, the medium is exchanged for phenol red-free medium and $25\ \si{\nano\mol}$ Hoechst 33342 (Invitrogen) is added to stain the nuclei.

\subsection{Tracking} \label{app:tracking}
To obtain the cell trajectories, we use the frames of DAPI-stained cell nuclei of individual cells and track them using LIM tracker \cite{aragaki2022lim}. We use a segmentation algorithm inside LIM tracker to detect cells in individual frames, checking every frame and manually correcting if necessary. The cell position is defined as the center of mass of the segmented area. Tracking is done using the built-in tracking functionality, but each track of each cell is checked and corrected if needed. This way, we analyze $17$ different experiments for MDA-MB-231 and MCF10A cells, and $14$ experiments for HT1080 cells. Representative trajectories are shown in Fig.~\ref{Fig1}(e)-(g).

\subsection{Behavior statistics} \label{app:methods_behavior_statistics}

To quantify collective cell behavior, we compute four statistics from the tracked cell positions $x_i$ and velocities $v_i$ (Methods~\ref{app:tracking}). Velocities are estimated by finite differences with a measurement interval of $\Delta t = 10\,\si{\minute}$. All statistics are conditioned on the packing density $\phi$, using bins of width $2/N_\phi$, where $N_\phi = L/\langle \ell \rangle$ is the number of cells needed to cover a track of length $L$ and $\langle \ell \rangle$ is the mean cell length at low density (Supplementary Section 6.1).

First, we quantify velocity coordination using the normalized velocity cross-correlation function $C(|x_{ij}|;\phi) = \langle v_i v_j \rangle_{\phi,|x_{ij}|}/\langle |v| \rangle_\phi^2$, where $x_{ij}=x_i-x_j$ is the relative cell position, $\langle |v| \rangle_\phi$ is the mean speed, and the numerator is averaged over cell pairs within bins of intercellular distance and density. Second, we characterize spatial organization using the pair correlation function $g(|x_{ij}|;\phi) = p(|x_{ij}|;\phi)/p_0(|x_{ij}|)$, where $p$ is the measured distribution of pairwise distances and $p_0(|x_{ij}|) = 2/L - 2|x_{ij}|/L^2$ is the corresponding distribution for uniformly distributed points on a line. Third, we calculate the exchange frequency $f_\mathrm{ex} = N_\mathrm{ex}/T_\mathrm{prox}$ as the number of cell exchanges normalized by the total time that cell pairs spend in close proximity ($|x_{ij}| < 30\,\si{\micro\metre}$).

Finally, we identify groups of cells using DBSCAN (Supplementary Section 5.1), with a neighborhood radius of $60\,\si{\micro\metre}$ and a minimum of three neighboring cells. Each group is classified according to its translational order $T = \left|N^{-1}\sum_i v_i/|v_i|\right|$: groups with $T>0.5$ are flocks and those with $T<0.5$ are clusters. Cells outside these groups are classified as individuals. We calculate the fraction of cells in each behavior category over all frames and experimental realizations at each density. The resulting statistics are robust to the choice of classification parameters (Supplementary Section 5.3). For more details refer to Supplementary Section 5.1.

\subsection{Single cell dynamics}\label{app:methods_basis}
We describe the projected nucleus positions $x_i(t)$ of all cells on a track by Eq.~(\ref{eq:GLE}). The single-cell term $F(x_i,v_i)$ is generally written as
\begin{equation}\label{eq_single_cell_basis_mt}
    F(x_i,v_i) = \gamma v_i + \sum_{n = 1} \Gamma_{n} v_i^{2n+1} + \epsilon \left(\frac{x_i -c}{c}\right)^z. 
\end{equation}
Here, $\gamma$ represents an effective friction coefficient, $\Gamma_{n}$ coefficients for higher order terms, and $\epsilon$ the strength of the confining potential. The confining potential is centered around $c=L/2$, with $L = 500 \, \si{\micro\meter}$ being the length of the micropatterned track, and the steepness of the potential is set to $z=11$. In Supplementary Section 3.3, we infer this equation of motion from single-cell experiments, through which we determine $\gamma$, $\Gamma_{n}$, and $\epsilon$ (Supplementary Table S1). By validating the inferred models on autocorrelation functions and velocity distributions, we show that higher order single cell terms are not needed. Thus, we use
$F(x_i,v_i)=\gamma v_i+\epsilon[(x_i-c)/c]^{z}$ consisting of an effective linear friction and a soft confinement at the track ends. Note that in the multicellular inference, $\gamma$ and $\epsilon$ are inferred again independently from the data. The inferred parameters do not vary greatly between single cell and multicellular experiments (Supplementary Table S1). Finally, the noise term in Eq.~(\ref{eq:GLE}) is inferred from data as well. We show that additive noise is enough and no multiplicative noise is needed to capture the single cell dynamics (Supplementary Section 3.3). For more details refer to Supplementary Sections 1.1 and 3.

\subsection{Memory}\label{app:methods_memory}
We write the memory term in Eq.~(\ref{eq:GLE}) as $M[v_i](t)=\beta\int_{-\infty}^{t}\d t'\,e^{-(t-t')/\tau}v_i(t')$ (Supplementary Section 1.1.3). This form has been shown to capture the long-time velocity autocorrelation of migrating cells \cite{selmeczi2005cell}. Because this non-Markovian term cannot be inferred directly, we determine $\beta$ and $\tau$ by matching simulated and experimental velocity autocorrelation functions (Supplementary Section 3.2). The memory parameters are found for both the single-cell experiments as well as for the multicellular experiments (Supplementary Table S1, Supplementary Fig. S4). Our interaction inference is largely robust against the choice of memory parameters (Supplementary Section 9.2). To correct the Markovian components in Eq.~(\ref{eq:GLE}) in the presence of the memory term, we develop a memory corrected inference procedure in Supplementary Section  2. This inference procedure is validated on simulated data (Supplementary Section 2.2).

\subsection{Interactions}
For each interaction class, we expand $G$ as in Eq.~(\ref{eq_expansion}), using $N_\mathrm{exp}$ exponentials with fixed decay lengths for the kernels and the odd terms of an $n$th-order polynomial for the directional dependencies (Supplementary Section 1.1.2). To avoid unconstrained extrapolation into rarely sampled high-velocity regions, all interaction basis functions are multiplied by a smooth velocity cutoff (Supplementary Section 1.3). In addition, we test different bases for the radial interaction kernel $g_\alpha$, but find that our inference is robust with respect to the choice of basis (Supplementary Section 9.3).

\subsection{Inference and basis reduction}\label{app:methods_inference}
We infer the coefficients of $F$ and $G$ and the noise amplitude $\sigma$ with Underdamped Langevin Inference \cite{bruckner2020inferring}, which estimates velocities and accelerations from positions sampled at $\Delta t=10\,\si{\minute}$ and corrects for biases due to discrete sampling and measurement noise (Supplementary Section 1.2). We extend the ULI force estimator with a correction for the memory term (Supplementary Section 2). To remove basis functions that are not resolved by the data, we infer the coefficients on $128$ bootstrap resamplings of whole experiments, remove the basis function with the smallest ratio $|\langle c\rangle|/\Delta c$ of mean to bootstrap standard deviation if this ratio is below $3$, and repeat until all remaining coefficients exceed this threshold (Supplementary Section 1.1.4). The final number of retained interaction basis functions $N_b$ defines the complexity of a model. For more details refer to Supplementary Section 1.

\subsection{Simulations}\label{app:methods_simulations}
We simulate the inferred models with an Euler--Maruyama scheme with a time step of $1\,\si{\minute}$ and record positions every $10\,\si{\minute}$, as in the experiments. To sample a range of densities, simulations start from cell numbers representative of the experiments and include stochastic cell division at a rate matched to the increase in cell number in the experiments. For more details refer to Supplementary Section 1.4.

\subsection{Statistical analysis}\label{app:methods_statistics}
Experimental behavior statistics (velocity correlation, pair correlation, exchange frequency, and flock and cluster fractions) are computed in packing-density bins (Methods~\ref{app:methods_behavior_statistics}). The sizes of the experimental datasets used for the inference and model selection are shown in Table S3. Experimental variability of the behavior statistics is quantified by $B=128$ bootstrap resamplings of whole experiments ($N=17$ experiments for MCF10A and MDA-MB-231 cells and $N=14$ for HT1080 cells). Across these bootstrapped samples, we compute averages and standard error of the mean (corresponding to 68\% bootstrap percentile CI, approximately equal to 1$\sigma$) as shown in Fig.~\ref{Fig2}. To quantify the variability of the behavior statistics predicted from the inferred models, we conduct our inference procedure $R = 128$ times on randomly chosen subsets of the short-time scale acceleration data with replacement. This yields $R = 128$ slightly different inferred models, each of which we simulate to predict the multicellular dynamics and the corresponding behavior statistics. The average of these simulations yields the average simulation behavior statistics and the standard deviation represents the standard error of the mean of the simulation behavior statistics as shown in Fig.~\ref{Fig2}. The spread of the inferred interaction kernels is shown in Fig.~\ref{Fig4}, with the shaded area indicating the 68\% bootstrap percentile CI, approximately equal to 1$\sigma$. 

\subsection{Model validation and selection}\label{app:methods_model_selection}
For model validation and selection, we combine computed behavior statistics into vectors for each experimental and model bootstrap sample. Following the statistical sampling procedures described in Methods section~\ref{app:methods_statistics}, we obtain distributions of $B=128$ experimental vectors $s_{\mathrm e}^{(b)}$ and $R=128$ simulated vectors $s_m^{(r)}$ for each candidate model $m$. To compare these vectors, we define a weighted Euclidean distance $d(s,s')$, in which each coordinate is standardized by its experimental bootstrap mean and standard deviation, and weights balance the contributions of the different behavior statistics (Supplementary Section 4.2). For model selection, we quantify the agreement between experimental and simulated behavior statistics using the energy distance~\cite{rizzo2016energy},
\begin{equation}
\begin{aligned}
D_{\mathrm E}(s_{\mathrm e},s_m)
&=\frac{2}{BR}\sum_{b=1}^{B}\sum_{r=1}^{R}
d(s_{\mathrm e}^{(b)},s_m^{(r)})\\
&-\frac{1}{B^2}\sum_{b,b'=1}^{B}
d(s_{\mathrm e}^{(b)},s_{\mathrm e}^{(b')})\\
&-\frac{1}{R^2}\sum_{r,r'=1}^{R}
d(s_m^{(r)},s_m^{(r')}).
\end{aligned}
\end{equation}
We quantify model complexity by the number of retained basis functions $N_{\mathrm b}$ and construct a complexity front by retaining, for each $N_{\mathrm b}$, the model with the lowest energy distance, provided it outperforms all simpler models. We then compare each model $m$ on this front against the model $m^*$ with the overall lowest energy distance, using the paired differences
\begin{equation}
\Delta D_{\mathrm E}^{(b)}
=D_{\mathrm E}(s_{\mathrm e}^{(b)},s_m)
-D_{\mathrm E}(s_{\mathrm e}^{(b)},s_{m^*}),
\end{equation}
where the energy distance between a single experimental realization $b$ and the simulated distribution of model $m$ is defined as
\begin{equation}
D_{\mathrm E}(s_{\mathrm e}^{(b)},s_m)
=\frac{2}{R}\sum_{r=1}^{R}d(s_{\mathrm e}^{(b)},s_m^{(r)})
-\frac{1}{R^2}\sum_{r,r'=1}^{R}d(s_m^{(r)},s_m^{(r')}).
\end{equation}
We select the model with the fewest basis functions satisfying
\begin{equation}
q_{0.05}\left(\{\Delta D_{\mathrm E}^{(b)}\}_{b=1}^{B}\right)\leq 0,
\end{equation}
where $q_{0.05}$ denotes the 5th percentile of the paired differences across experimental bootstrap realizations. Further details of the model selection procedure are provided in Supplementary Section 4.

\section*{Data availability}
The cell trajectories used in this study are available upon request.

\section*{Code availability}
The code for inference, simulation, and model selection will be made available at a public repository.

\bibliography{final_references}

\end{document}